\pdfoutput=1
\documentclass[aps,prl,reprint,groupedaddress]{revtex4-1}

\usepackage{amssymb}
\usepackage{amsmath}
\usepackage[colorlinks=true,linkcolor=blue,citecolor=blue, urlcolor=blue]{hyperref}   
\usepackage{tikz}
\usepackage{graphicx}
\usepackage[utf8]{inputenc}
\newcommand{\kompost}{K{\o}MP{\o}ST}

\definecolor{uibred}{RGB}{167, 38, 47}
\def\Eq#1{Eq.~(\ref{#1})}

\def\Fig#1{Fig.~\ref{#1}}

\def\x{\mathbf{x}}
\newcommand{\xt}{\mathbf{x}}

\newcommand{\TBg}{\overline{T}}
\newcommand{\TId}{T_\text{id}}

\newcommand{\tauekt}{\tau_\textsc{ekt}}
\newcommand{\tauhydro}{\tau_\text{hydro}}
\def\st{\begin{equation}}
\def\stp{\end{equation}}
\def\llangle{\left\langle}
\def\rrangle{\right\rangle}
\usepackage{tikz}

\begin{document}

\title{Matching the Nonequilibrium Initial Stage of Heavy Ion Collisions to Hydrodynamics with QCD Kinetic Theory}

\author{Aleksi Kurkela}
\email[]{a.k@cern.ch}
\affiliation{Theoretical Physics Department, CERN, Geneva, Switzerland}
\affiliation{Faculty of Science and Technology, University of Stavanger, 
4036 
Stavanger, Norway}

\author{Aleksas Mazeliauskas}
\email[]{a.mazeliauskas@thphys.uni-heidelberg.de}
\affiliation{Institut f\"{u}r Theoretische Physik, Universit\"{a}t Heidelberg, 
69120 Heidelberg, Germany}
\affiliation{Department of Physics and Astronomy, Stony Brook University, Stony 
	Brook, New York 11794, USA}

\author{Jean-Fran\c cois Paquet}
\email[]{jeanfrancois.paquet@duke.edu}
\affiliation{Department of Physics, Duke University, Durham, North Carolina 27708, USA}
\affiliation{Department of Physics and Astronomy, Stony Brook University, Stony 
Brook, New York 11794, USA}

\author{S\"{o}ren Schlichting}
\email[]{sschlichting@physik.uni-bielefeld.de}
\affiliation{Fakult\"{a}t f\"{u}r Physik, Universit\"{a}t Bielefeld, D-33615 Bielefeld, Germany}
\affiliation{Department of Physics, University of Washington, Seattle, Washington 
98195-1560, USA}

\author{Derek Teaney}
\email[]{derek.teaney@stonybrook.edu}
\affiliation{Department of Physics and Astronomy, Stony Brook University, Stony 
	Brook, New York 11794, USA}

\date{\today}

\begin{abstract}

High-energy nuclear collisions produce a nonequilibrium plasma of quarks and gluons which thermalizes and exhibits
hydrodynamic flow. 
There are currently no practical frameworks to connect
the early particle production in classical field simulations to the subsequent
hydrodynamic evolution. We build such a framework using
nonequilibrium Green's functions, calculated in QCD kinetic theory, to
propagate the initial energy-momentum tensor to the hydrodynamic phase.
We demonstrate that this approach can be easily incorporated into
existing hydrodynamic
simulations, leading to stronger constraints on the energy density at early times and the transport properties of the QCD medium.
Based on (conformal) scaling properties of the Green's functions, we further obtain pragmatic bounds for the applicability of hydrodynamics in nuclear collisions.

\end{abstract}

\maketitle

Collisions of heavy nuclei at the Relativistic
Heavy Ion Collider (RHIC) and the Large Hadron Collider (LHC) heat up nuclear
matter sufficiently to produce a plasma of deconfined colored
degrees of freedom ---  the quark-gluon plasma (QGP)~\cite{Adams:2005dq,Adcox:2004mh,Back:2004je,Arsene:2004fa}. 
The properties of this deconfined plasma can only be constrained indirectly from the mass and momentum distribution of the final shower of color-neutral particles reaching the detectors.
Extensive comparisons with measurements indicate that the spacetime
evolution of the QGP can be described with relativistic viscous
hydrodynamics~\cite{Heinz:2013th,Teaney:2009qa,Luzum:2013yya,Gale:2013da,deSouza:2015ena} starting from a time $\tau \equiv \tauhydro \sim 1$~fm/c after
the collision, with other models describing the dynamics of the collision before and after this hydrodynamic phase. 
The success of hydrodynamic models
has made it possible to study certain finite
temperature properties of quantum chromodynamics (QCD), such as the specific shear viscosity $\eta/s$. Most remarkably it
was found that $\eta/s$ is on the order of a tenth in units of $\hbar/k_B$
(with $k_B$ being the Boltzmann constant), perhaps the smallest specific shear
viscosity  ever measured. 
Because constraints on the plasma's properties are obtained from multistage simulations, an intricate interdependency exists between the hydrodynamic evolution and the description of the earlier phase of
heavy ion collisions.
Obtaining  more precise constraints on $\eta/s$ and
other transport coefficients of QCD 
hinges to a
considerable extent on attaining a better understanding of this early stage of
the collisions and its transition to hydrodynamics. 

Before the time $\tauhydro$, the deconfined matter is not amenable to a coarse-grained description in terms of macroscopic hydrodynamics fields, and a microscopic description must be used.
Modeling the earliest phase of heavy ion collisions and its transition to hydrodynamics from first-principles QCD remains a formidable challenge. In this Letter, we address this challenge by showing how an effective kinetic theory (EKT) of weakly coupled QCD~\cite{Arnold:2002zm} can be used to smoothly describe the evolution of a general out-of-equilibrium energy-momentum tensor specified at very early time $\tauekt \ll \tauhydro$ to its late time hydrodynamic form.

Naturally, our weakly coupled approach is particularly well suited for collisions of large nuclei at high energies, where a quantitative theory of gluon production can be developed based on the color glass condensate theory of high-energy QCD \cite{Iancu:2002xk,Iancu:2003xm,Gelis:2010nm,Gelis:2007kn}. Based on a separation into slow and fast degrees of freedom, the early time dynamics of the collision in the color glass condensate approach is described in terms of classical gluon fields. Classical Yang-Mills equations determine the dynamics of the system until a time $\tauekt$ at which the phase space density of gluons (per $\hbar$) becomes less than $\sim 1/\alpha_s(Q_s)$, with $\alpha_s$ being the strong coupling constant evaluated at the saturation scale $Q_s$. Subsequently, for $\tau>\tauekt$ kinetic processes dominate the spacetime evolution of the highly occupied and highly anisotropic plasma of gluons~\cite{Berges:2013fga,Berges:2014yta}; detailed simulations of the classical Yang-Mills dynamics~\cite{Berges:2013eia,Berges:2013fga,Berges:2014yta} confirm the onset of the ``bottom-up" thermalization scenario developed in a seminal paper many years ago~\cite{Baier:2000sb}. Since then there have been a multitude of analytical and numerical works devoted to clarifying the non-Abelian field dynamics in the earliest stages of the evolution~\cite{Krasnitz:2003jw,Lappi:2006fp,Lappi:2011ju,Epelbaum:2013waa,Berges:2013eia, Berges:2013fga, Gelis:2013rba,Berges:2014yta,Schenke:2015aqa}, and to developing tools to simulate the subsequent approach towards local thermal equilibrium using QCD kinetics~\cite{Xu:2004mz,El:2007vg,Greif:2017bnr,York:2014wja,Kurkela:2014tea,Kurkela:2015qoa,Keegan:2016cpi}, involving several unique features such as non-Abelian collinear radiation~\cite{Baier:1996kr} and dynamical screening~\cite{Arnold:2002zm}. In spite of this theoretical progress, a practical tool to bridge between the early time dynamics of strong color fields and successful hydrodynamic simulations at late times has not been achieved so far.

Based on a nonequilibrium linear response formalism~\cite{Keegan:2016cpi} developed in detail in our
companion paper~\cite{Kurkela:2018vqr}, our Letter provides for the
first time a concrete realization of a satisfactory theoretical description of
the early time out-of-equilibrium dynamics. The computer code, called \kompost{} after its authors, is publicly available~\cite{kompost_github}. 
While the underlying kinetic approach can  be justified rigorously
for the collision of large nuclei only in the limit of very weak coupling,
we show that 
the kinetic response to a variety of initial conditions 
can be smoothly extrapolated to physically relevant 
couplings by using an appropriate scaling variable.
This makes \kompost{} a practical tool for RHIC and LHC phenomenology.

\begin{figure}
	\centering
\includegraphics[width=\linewidth]{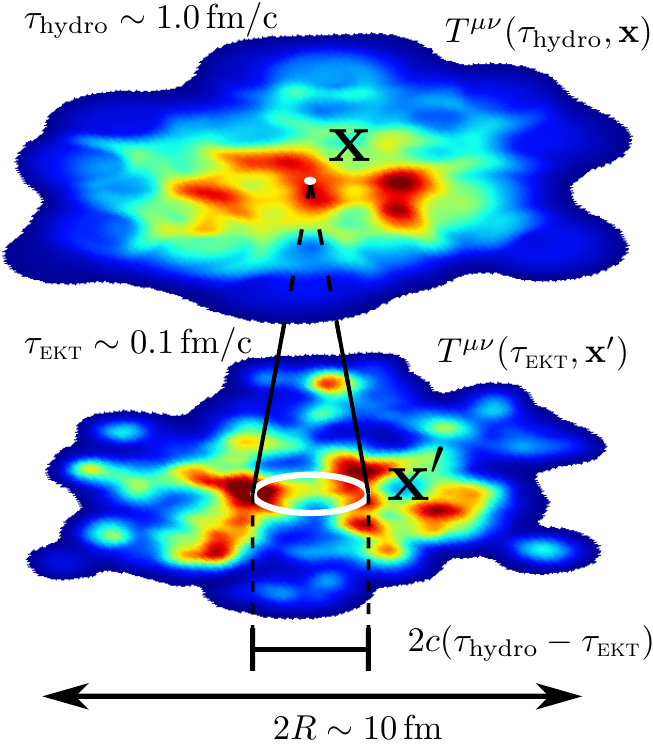}
	\caption{\label{fig:glasma3d} Evolution of the transverse energy density profile calculated within the linear response framework \kompost{},  from early time $\tauekt\sim 0.1\,\text{fm/c}$ to hydrodynamization time $\tauhydro\sim 1.0\,\text{fm/c}$. The energy-momentum tensor at each point $\x$ in the transverse plane receives causal contributions from the local average background,
		and linearized energy and momentum perturbations propagated from $\tauekt$ to $\tauhydro$ 
		(\Eq{eq:pert_evol}). The causal past for point $\x$ is indicated by the black cone and white circle.}
\end{figure}

\noindent\emph{Nonequilibrium linear response theory.}---  
Using a nonequilibrium linear response formalism, the
energy-momentum tensor of the system is evolved from its nonequilibrium
form at $\tauekt$ up to a time $\tauhydro$ when hydrodynamics
becomes applicable -- see \Fig{fig:glasma3d}.
Our approximation scheme is based on the  separation of scales seen in
\Fig{fig:glasma3d}~\cite{Keegan:2016cpi}: to determine
the energy density at a spacetime point ($\tauhydro$,$\x$), one  needs only
to propagate the nonequilibrium initial condition from points ($\tauekt$,$\x'$) in the causal past
 $|\x -\x'|  <
c(\tauhydro-\tauekt)$. Since this causal circle is small compared
to the system size $\sim 2  R$,  the energy-momentum tensor in this domain can be
divided into a local average and (small) perturbations~\footnote{Note that in small collision systems such as $p$+$p$ or $p$+Pb, a separation of scales is not warranted, and the linearized description underlying \kompost{} may be inapplicable}
\begin{equation}
T^{\mu\nu}(\tauekt,\x')=\TBg^{\mu\nu}_\x(\tauekt)+\delta T^{\mu\nu}_\x(\tauekt,\x').
\label{eq:Tmunu_decomp}
\end{equation}
In practice, the background $\TBg^{\mu\nu}_\x(\tau)$ is calculated 
by a spatial average of $T^{\mu\nu}(\tauekt,\x')$ over the
causal circle, and is assumed to be  boost invariant and locally homogeneous in the transverse ($xy$-)plane. Differences between the full energy-momentum tensor  $T^{\mu\nu}(\tauekt,\x')$ and the background $\TBg^{\mu\nu}_\x(\tauekt)$ are treated as linearized perturbations $\delta T^{\mu\nu}_\x(\tauekt,\x')$, which propagate according to
\begin{align}
	\delta 
	T^{\mu\nu}_\x(\tauhydro,\x) &=\int 
	d^2\xt'~G^{\mu\nu}_{\alpha 
		\beta}\left(\xt,\xt',\tauhydro,\tauekt\right)\nonumber\\
&\times\delta 
	T_\x^{\alpha\beta}(\tauekt,\xt') \frac{\TBg^{\tau \tau}_\x(\tauhydro)}{\TBg^{\tau\tau}_\x(\tauekt)}\label{eq:pert_evol}.
\end{align}
Here the Green's functions $G^{\mu\nu}_{\alpha
\beta}\left(\xt,\xt',\tauekt,\tauhydro\right)$ describe the evolution and
equilibration of perturbations from an early time $\tauekt$ to a later time
$\tauhydro$.

Notably this linear response formalism provides a general framework to calculate the evolution of the energy-momentum tensor, which requires limited microscopic input in the form of the nonequilibrium evolution of the background and linearized response functions. In this Letter, this microscopic input is calculated in QCD kinetic theory~\cite{Arnold:2002zm}.
Starting from a microscopic gluon distribution function motivated by classical simulations of early time
dynamics~\cite{Berges:2013eia}, we numerically solve the Boltzmann equation with the pure-glue leading
order QCD collision kernels~\cite{Arnold:2002zm,Kurkela:2015qoa,Keegan:2016cpi,Kurkela:2018vqr}. 
By analyzing the time dependence of the background distribution function and its perturbations, the evolution of the background energy-momentum tensor and the Green's functions are then extracted from the kinetic theory simulations~\cite{Kurkela:2018vqr}. 

\noindent\emph{Equilibration time and conformal scaling.}--- 
Based on the approximate conformal symmetry of high-energy QCD, 
the rate of equilibration of the background energy-momentum tensor $\TBg^{\mu\nu}$ is governed solely by the QCD coupling constant $\lambda=4\pi\alpha_s N_c$ along with a single dimensionful scale parameter. Since kinetic theory approaches hydrodynamics at late time, it is natural to express the dimensionful scale in terms of an asymptotic equilibrium quantity. Defining a pseudo-temperature $\TId \equiv (\tau^{1/3} T)_\infty / \tau^{1/3} $ based on the asymptotic $\tau^{-1/3}$ dependence of the temperature in (conformal) Bjorken hydrodynamics, the equilibration rate is then determined by the kinetic relaxation time $\tau_\text{R}(\tau) \equiv (\eta/s)/\TId(\tau)$, where the dependence on the coupling constant $\lambda$ is encoded in the specific shear viscosity $\eta/s$.

\begin{figure}
	\centering
	\includegraphics[width=\linewidth]{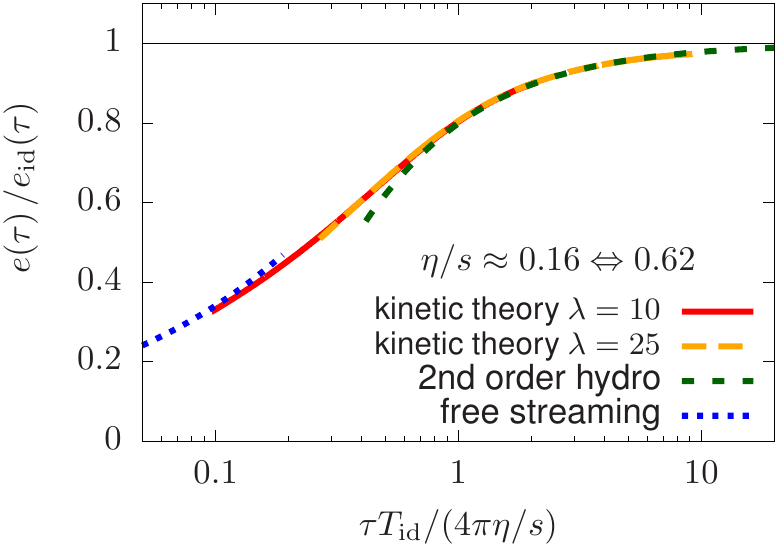}
	\caption{Evolution of the background energy density in kinetic theory for two values of the coupling constant $\lambda$, corresponding to a range of specific shear viscosities $\eta/s\approx 0.16{-}0.62$. Scaling the vertical axis by the ideal hydrodynamic asymptotics 
		$e_\text{id.}=\nu_g \pi^2\TId^4/30$ 
		and the horizontal axis by the kinetic relaxation time $\tau_\text{R}(\tau) \equiv (\eta/s)/\TId(\tau)$ reveals that the nonequilibrium evolution follows a universal attractor curve which smoothly interpolates between free streaming at early times and viscous hydrodynamics at late times.}
	\label{fig:scaling}
\end{figure}

In contrast to earlier expectations~\cite{Baier:2000sb} based on parametric estimates at (asymptotically) weak coupling, we find that for moderate values of 
$\eta/s$ $(\lesssim 1)$ the timescale $\tauhydro$ on which the nonequilibrium plasma approaches (viscous) hydrodynamics is determined by the equilibrium relaxation rate $\tau_{R}$~\cite{Heller:2016rtz,Kurkela:2015qoa}. Specifically, we observed in our kinetic theory simulations that the nonequilibrium evolution of the background energy density exhibits a universal scaling behavior. Expressing the evolution time $\tau$ in units of the relaxation time $\tau_R$,  the ratio of the energy density $e(\tau)$ to the asymptotic ideal hydrodynamic value $e_\text{id.}\sim \TId^4$, shown in \Fig{fig:scaling}, becomes independent of the microscopic coupling strength and smoothly interpolates between an approximate free-streaming behavior at early times and a hydrodynamic evolution at late times.

Based on the assumed symmetries, the universal curve in \Fig{fig:scaling} determines the equilibration of the entire background energy-momentum tensor $\TBg^{\mu\nu}(\tau)$. Such a collapse of the macroscopic evolution has also been referred to as a ``hydrodynamic attractor"  and is actively studied in the literature for different microscopic descriptions (mostly) in boost-invariant conformal systems~\cite{Heller:2016rtz,Romatschke:2017vte,Strickland:2017kux,Behtash:2017wqg}.
Crucially, we found that the linear kinetic response functions $G^{\mu\nu}_{\alpha\beta}(\x,\x', \tauekt, \tauhydro)$  used to propagate initial energy and momentum perturbations in \Eq{eq:pert_evol} do not depend separately on the evolution time, background energy density, or the coupling constant, but only through the ratio of $\tau/\tau_R$.
One important consequence of this result is that the response functions  need to be evaluated only once in a full kinetic theory simulation and can be reused  for a different value of $\eta/s$ or background temperature scale $\TId(\tau)$ by simple scaling transformations~\cite{Kurkela:2018vqr}.

Besides its practical utility, the universality of the equilibration process in kinetic theory enables us to make pragmatic estimates of the time necessary before the plasma created in high-energy collisions can be described with hydrodynamics. We infer from \Fig{fig:scaling} that the kinetic theory evolution approaches the hydrodynamic limit on time scales $\tauhydro \approx 4\pi \tau_\text{R}$. Relating the asymptotic constant $ (\tau^{1/3} T)_\infty$ to the average entropy density per unit rapidity $\llangle \tau s\rrangle$,
we obtain the following estimate for the hydrodynamization time
\begin{equation}
\label{eq:hydrotime}
\tauhydro\approx 1.1\,{\rm fm} \, \left( \frac{4\pi(\eta/s)}{2} \right)^{{3}/{2}}  \left( \frac{ \llangle \tau s\rrangle }{ 4.1 \, {\rm  GeV}^2 } \right)^{-1/2} \left( \frac{\nu_\text{eff}}{40} \right)^{1/2}.
\end{equation}
Here $\nu_\text{eff}=\nu_{\rm eff}(T_{\rm eq})$ is the effective number of
degrees of freedom in the plasma at the equilibration temperature, as
determined from the equilibrium relation $s(T)=\nu_\text{eff}
\frac{4\pi^2}{90}T^3$.  We use $\nu_g=2(N_c^2-1)=16$ in our numerical simulations of a gluonic plasma,
while for a realistic QCD equation of state, $\nu_{\rm
eff}(0.4\,\text{GeV})\approx 40$~\cite{Bazavov:2014pvz,Borsanyi:2016ksw}. 
$\llangle \tau s\rrangle$ for PbPb collisions $\sqrt{s_{NN}}=2.76\,\text{TeV}$ is tightly constrained by experimental measurements  and
hydrodynamic simulations  to be
$\llangle \tau s\rrangle\approx 4.1 \, {\rm  GeV}^2 $~\cite{Keegan:2016cpi}.

\noindent \emph{Dynamical description of pre-equilibrium stage:}---We now present results obtained by applying our kinetic propagator \kompost{} to a realistic boost-invariant initial conditions of a central Pb-Pb collision at a center-of-mass energy $\sqrt{s_{NN}}=2.76\,\text{TeV}$. We start from an initial energy-momentum profile at $\tauekt=0.2\,\text{fm}$~\footnote{Small scale fluctuations ---less than $\tauekt$---were smeared from the IP-glasma  initial conditions to limit the strain on the linearized perturbation approximation.} given by
the impact parameter dependent (IP)-glasma model~\cite{Schenke:2012wb,Schenke:2012fw}, which provides a microscopic description of the classical Yang-Mills dynamics before the onset of equilibration. Subsequently, the locally averaged background energy-momentum tensor $\TBg^{\mu\nu}$ for each point in the transverse plane is evolved in kinetic theory according to
the universal evolution curve (see \Fig{fig:scaling}). Similarly, using appropriately scaled nonequilibrium response functions, initial energy ($\delta T^{\tau\tau}$), and transverse momentum perturbations ($\delta T^{\tau i}$) are propagated according to \Eq{eq:pert_evol} up to a time $\tauhydro$, which is varied around the estimate $\tauhydro\approx 0.6\,\text{fm}$,  evaluated according to \Eq{eq:hydrotime} with $\nu_g=16$ and $\langle s \tau \rangle\approx 5.0\,\text{GeV}^2$ for this particular event~\footnote{We neglect perturbations in the $\delta T^{ij}$, $\delta T^{\eta i}$, and $\delta T^{\eta\eta}$ components, which are typically small and get further washed out during the nonequilibrium evolution, as they are not related to conserved quantities.}. Beyond $\tauhydro$ the evolution is modeled using relativistic viscous hydrodynamics~\cite{Schenke:2010nt,Schenke:2010rr,Paquet:2015lta} with constant $\eta/s=2/(4\pi)$ and QCD equation of state~\cite{Huovinen:2009yb}.

\begin{figure}
	\includegraphics{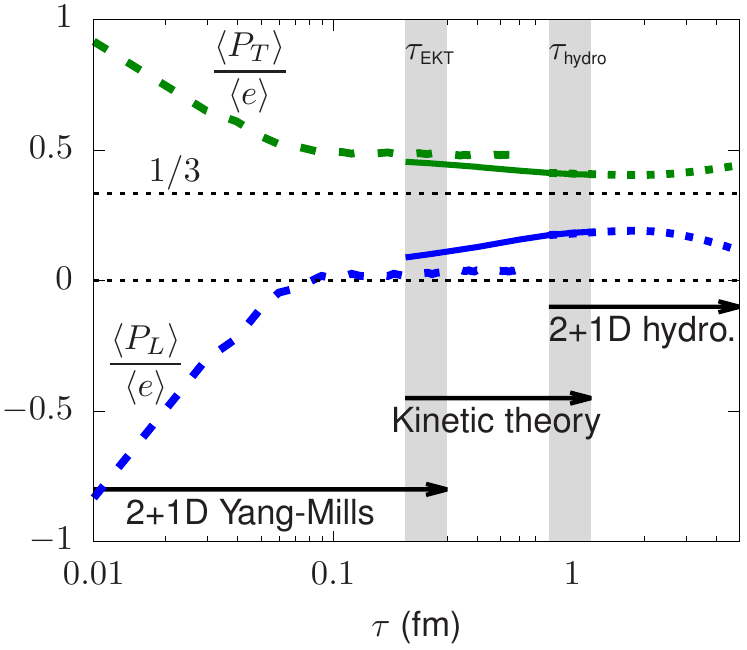}
	\caption{
		Average transverse and longitudinal pressures, $P_T= (T^{xx}+T^{yy})/2$ and $P_L=\tau^2 T^{\eta\eta}$, of a realistic heavy ion event evolved in succession by 2+1D Yang-Mills evolution (IP-glasma model)~\cite{Schenke:2012wb,Schenke:2012fw}, QCD kinetic theory (\kompost) and relativistic viscous hydrodynamics~\cite{Schenke:2010nt,Schenke:2010rr,Paquet:2015lta}.\label{fig:ipglasma-matching}}
\end{figure}

In \Fig{fig:ipglasma-matching} we show overlapping theoretical descriptions of the evolution of the pressure anisotropy in the early stages of a realistic event. In the classical Yang-Mills field simulations the longitudinal pressure  $\llangle P_L\rrangle$ is initially negative, as is typical of a classical field configuration (cf.\ parallel plate capacitor with electric field $E$, where $T^{ij} ={\rm diag} (E^2, E^2, -E^2)/2$~\cite{Jackson:1998nia}). As the system evolves, the classical fields lose coherence and the longitudinal pressure approaches zero;
the increasingly dilute system is then better  described by kinetic theory.  
In the kinetic phase ($\tauekt \rightarrow \tauhydro$) the energy-momentum tensor begins to equilibrate, such that ultimately
the pressure approaches its equilibrium value of $1/3$ of the energy density (for a locally equilibrated fluid of massless particles), up to corrections captured by viscous hydrodynamics. One clearly observes from \Fig{fig:ipglasma-matching} that the kinetic equilibration stage provides the missing link between the classical Yang-Mills evolution and the hydrodynamics, thus creating a self-consistent description of initial stages.  

\begin{figure}
\includegraphics{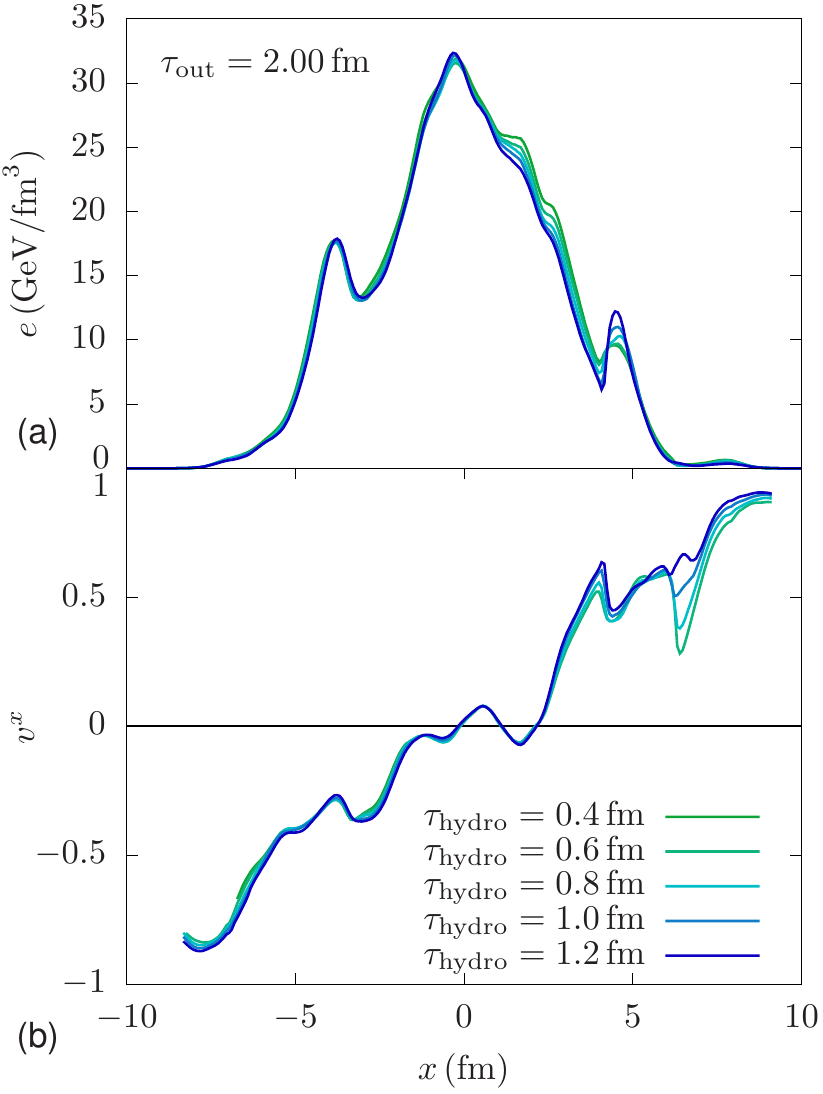}
	\caption{
		(a) Energy density and (b) velocity profiles in the hydrodynamic stage at time $\tau_\text{out}=2.0\,\text{fm}$, for different durations of the kinetic preequilibrium stage ($\tauekt \rightarrow \tauhydro$). \label{fig:ipglasma-profiles}
}
\end{figure}

In order to illustrate the quality of the matching between the kinetic theory and the hydrodynamics,
we investigate the robustness of the transverse energy and velocity profiles at $\tau_{\rm out}=2.0\,{\rm fm}$ 
(in the hydrodynamic stage) under variations of the duration of the preequilibrium evolution $\tauhydro$.
In \Fig{fig:ipglasma-profiles}(a)  we see that the energy density at  $\tau_{\rm out}$ is essentially unchanged as the initialization time $\tauhydro$ is varied,  indicating a smooth matching between the kinetic and hydrodynamic simulations. Figure \ref{fig:ipglasma-profiles}(b) shows the corresponding plot for the transverse velocities. The transverse flow is also smoothly matched between the kinetic and hydrodynamic phases, with tension  visible only at the edges of the fireball where our linearized approximation is pushed beyond its limits due to large gradients. We emphasize that \kompost{}  also provides the viscous stress tensor $\pi^{\mu\nu}$  which is required for the subsequent (viscous) hydrodynamic evolution. We verified that the obtained $\pi^{\mu\nu}$ agrees reasonably well with the Navier-Stokes constitutive equations ($\pi^{\mu\nu} \approx-\eta \sigma^{\mu\nu}$), guaranteeing a consistent matching between the kinetic and hydrodynamic simulations~\cite{Kurkela:2018vqr}. 

Our current approach using kinetic response functions should be compared to preequilibrium
modeling based on the long wavelength response~\cite{Vredevoogd:2008id,vanderSchee:2013pia,Romatschke:2015gxa} and free
streaming~\cite{Broniowski:2008qk,Liu:2015nwa}. 
We find that the first few terms in the long wavelength expansion capture most of the kinetic response~\cite{Kurkela:2018vqr}.  However, to achieve a satisfactory level of agreement one needs to go beyond the leading order velocity
response of Ref.~\cite{Vredevoogd:2008id}.
Because of the universality of the velocity response in conformal
systems~\cite{Vredevoogd:2008id,Keegan:2016cpi}, we find that the velocity is  well predicted even by free streaming. However,
in the absence of longitudinal pressure, the free-streaming energy density decreases significantly slower
than in kinetic theory or hydrodynamics. Moreover, the viscous stress tensor components $\pi^{\mu\nu}$ do not  approach their hydrodynamic limit in the free-streaming description compromising the smooth matching to the hydrodynamic phase.

\noindent \emph{Conclusions \& Outlook:}---Nonequilibrium linear response theory captures the essential features of the
early time preequilibrium dynamics of high-energy heavy ion collisions, and it
provides a practical framework  for connecting the theory of the initial state 
with the late time hydrodynamic evolution. Since the preequilibrium
dynamics from kinetic theory matches smoothly onto viscous hydrodynamics, the approach
naturally provides initial conditions for all second order hydrodynamic
variables. There is no need to readjust the energy density, flow velocity or shear stress
as the switching time $\tauhydro$ is changed, making the combined simulations increasingly predictive.
Hadronic observables such as the multiplicity $dN/d\eta$ or transverse momentum  $\llangle p_T \rrangle $
are now essentially insensitive to the switching time $\tauhydro$~\cite{Kurkela:2018vqr}.

Without the additional uncertainties from the initial state, 
near-thermal properties of the QCD medium
can be increasingly constrained through multistage simulations and it would be interesting to reperform statistical inferences of transport coefficients $\eta/s$ and $\zeta/s$~\cite{Bernhard:2016tnd} with \kompost{}. Conversely, the properties of the earliest stage of the collisions can also be better inferred from experiments. For example, the color glass condensate predictions for the multiplicity as a function of centrality~\footnote{See \cite{Aamodt:2010pb,Adam:2015ptt} and  references therein.} should be revisited to take into account the rapid production of entropy during the kinetic pre-equilibrium phase.

A further benefit of handling the rich preequilibrium dynamics within kinetic theory
is that
the physics of chemical equilibration, early parton energy loss and electromagnetic emission can all be naturally described within the same QCD kinetic framework. 
The future inclusion of quark degrees of freedom in our initial state propagator  \kompost{}~\cite{kompost_github} will be an important step in this direction.

Aside from improving the predictive power of state-of-the-art dynamical simulations of nucleus-nucleus collisions, the more complete understanding of the
early time dynamics can also be used to assess the limits of applicability of
the hydrodynamic description---a topic of considerable interest in light of recent
experimental results in proton-proton ($p$+$p$) and proton-nucleus ($p$+$A$) collision~\cite{Dusling:2015gta}. Since a
sufficiently long-lived hydrodynamic phase can  be realized only when the
equilibration time $\tauhydro$ is small compared to the system size $R$
\footnote{If this is not the case, the system rapidly expands in all three dimensions, resulting in a rapid decrease of the energy density such that the system freezes out before reaching the hydrodynamic stage.}, one can use the estimate in \Eq{eq:hydrotime} to
determine the smallest system that can hydrodynamize. Since the
entropy density per unit rapidity is approximately
conserved during the later stages of the evolution, we can
directly relate $\langle\tau s\rangle$ at equilibrium time to the charged particle multiplicity $dN_\text{\rm ch}/d\eta$ in the final state, according to
 $\langle\tau s\rangle \approx (S/N_\text{\rm ch})~1/A_\bot~dN_\text{\rm ch}/d\eta$, where the transverse area of the system is estimated by $A_{\bot}\approx \pi R^2$. Recalling that the ratio of entropy to
charged particle number is $S/N_\text{\rm ch}\approx 7$ for a hadron resonance gas~\cite{Muller:2005en,Hanusthesis}, we obtain that the ratio 
\begin{equation}
\label{eq:multestimate}
   \frac{\tauhydro}{R} \simeq \left( \frac{4\pi(\eta/s)}{2} \right)^{{3}/{2}}  \left( \frac{ dN_{\rm ch}/d\eta  }{63 } \right)^{-{1}/{2}}\!\!\left( \\
\frac{S/N_\text{ch}}{7} \right)^{-1/2}\!\! \left( \frac{\nu_{\rm eff}} {40}\right)^{{1}/{2}}
\end{equation}
is independent of the collision system size $R$ and  depends  only on the charged particle multiplicity $dN_\text{ch}/d\eta$ \cite{Basar:2013hea}
along with  equilibrium and transport properties of the QGP.
Then using \Eq{eq:multestimate} one concludes that in an optimistic scenario the minimum requirement for the hydrodynamic phase, i.e. $\tauhydro/R\approx 1$, is reached if the charged particle multiplicity is at least  $dN_{\rm ch}/d\eta \gtrsim 8$ for a small $\eta/s=1/(4\pi)$. However, this estimate is rather sensitive to the transport properties of the QGP as for a larger value of specific shear viscosity $\eta/s=2/(4\pi)$ the minimal required multiplicity increases to $dN_{\rm ch}/d\eta \gtrsim 63$. Based on the experimentally measured minimum bias multiplicities of $dN_\text{ch}/d\eta \simeq 6$ in  $7$~TeV $p$+$p$~\cite{Aamodt:2010pp} and $dN_\text{ch}/d\eta \simeq 16$ in $5.02$~TeV $p$+Pb collisions~\cite{ALICE:2012xs}, one concludes that a hydrodynamic phase is unlikely to emerge in most $p$+$p$ collisions. However in $p$+Pb collisions, where events with several (up to $\sim10$) times the minimum bias multiplicity can be observed, it seems possible that a hydrodynamically flowing QGP can be formed.

\begin{acknowledgments}

The authors would like to thank Bj\"orn Schenke for the insightful discussions and for his help adapting the hydrodynamics code \textsc{music} for this work, and
Liam Keegan for his contributions at the beginning of this project. Useful discussions with J\"urgen Berges, Stefan Fl\"orchinger, Yacine Mehtar-Tani, Klaus Reygers, and Raju Venugopalan are gratefully acknowledged.
Results in this paper were obtained using the high-performance computing system 
at the Institute for Advanced Computational Science at Stony Brook University. 
This work was supported in part by the U.S. Department of Energy, Office of 
Science, Office of Nuclear Physics  under Awards No. 
DE\nobreakdash-FG02\nobreakdash-88ER40388 (A.M., J.-F.P., D.T.), No. DE-FG02-05ER41367 (J.-F.P.), and No. 
DE-FG02-97ER41014 (S.S.). This work was supported in part by the German Research Foundation (DFG) 
Collaborative Research Centre (SFB) 1225 (ISOQUANT) (A.M.). Finally, A.M.
would like to thank the CERN Theoretical Physics
Department for the hospitality during the short-term visit.
\end{acknowledgments}

\bibliography{master.bib}

\begin{thebibliography}{66}%
\makeatletter
\providecommand \@ifxundefined [1]{%
 \@ifx{#1\undefined}
}%
\providecommand \@ifnum [1]{%
 \ifnum #1\expandafter \@firstoftwo
 \else \expandafter \@secondoftwo
 \fi
}%
\providecommand \@ifx [1]{%
 \ifx #1\expandafter \@firstoftwo
 \else \expandafter \@secondoftwo
 \fi
}%
\providecommand \natexlab [1]{#1}%
\providecommand \enquote  [1]{``#1''}%
\providecommand \bibnamefont  [1]{#1}%
\providecommand \bibfnamefont [1]{#1}%
\providecommand \citenamefont [1]{#1}%
\providecommand \href@noop [0]{\@secondoftwo}%
\providecommand \href [0]{\begingroup \@sanitize@url \@href}%
\providecommand \@href[1]{\@@startlink{#1}\@@href}%
\providecommand \@@href[1]{\endgroup#1\@@endlink}%
\providecommand \@sanitize@url [0]{\catcode `\\12\catcode `\$12\catcode
  `\&12\catcode `\#12\catcode `\^12\catcode `\_12\catcode `\%12\relax}%
\providecommand \@@startlink[1]{}%
\providecommand \@@endlink[0]{}%
\providecommand \url  [0]{\begingroup\@sanitize@url \@url }%
\providecommand \@url [1]{\endgroup\@href {#1}{\urlprefix }}%
\providecommand \urlprefix  [0]{URL }%
\providecommand \Eprint [0]{\href }%
\providecommand \doibase [0]{http://dx.doi.org/}%
\providecommand \selectlanguage [0]{\@gobble}%
\providecommand \bibinfo  [0]{\@secondoftwo}%
\providecommand \bibfield  [0]{\@secondoftwo}%
\providecommand \translation [1]{[#1]}%
\providecommand \BibitemOpen [0]{}%
\providecommand \bibitemStop [0]{}%
\providecommand \bibitemNoStop [0]{.\EOS\space}%
\providecommand \EOS [0]{\spacefactor3000\relax}%
\providecommand \BibitemShut  [1]{\csname bibitem#1\endcsname}%
\let\auto@bib@innerbib\@empty
\bibitem [{\citenamefont {Adams}\ \emph {et~al.}(2005)\citenamefont {Adams}
  \emph {et~al.}}]{Adams:2005dq}%
  \BibitemOpen
  \bibfield  {author} {\bibinfo {author} {\bibfnamefont {J.}~\bibnamefont
  {Adams}} \emph {et~al.} (\bibinfo {collaboration} {STAR}),\ }\href {\doibase
  10.1016/j.nuclphysa.2005.03.085} {\bibfield  {journal} {\bibinfo  {journal}
  {Nucl. Phys.}\ }\textbf {\bibinfo {volume} {A757}},\ \bibinfo {pages} {102}
  (\bibinfo {year} {2005})},\ \Eprint {http://arxiv.org/abs/nucl-ex/0501009}
  {arXiv:nucl-ex/0501009 [nucl-ex]} \BibitemShut {NoStop}%
\bibitem [{\citenamefont {Adcox}\ \emph {et~al.}(2005)\citenamefont {Adcox}
  \emph {et~al.}}]{Adcox:2004mh}%
  \BibitemOpen
  \bibfield  {author} {\bibinfo {author} {\bibfnamefont {K.}~\bibnamefont
  {Adcox}} \emph {et~al.} (\bibinfo {collaboration} {PHENIX}),\ }\href
  {\doibase 10.1016/j.nuclphysa.2005.03.086} {\bibfield  {journal} {\bibinfo
  {journal} {Nucl. Phys.}\ }\textbf {\bibinfo {volume} {A757}},\ \bibinfo
  {pages} {184} (\bibinfo {year} {2005})},\ \Eprint
  {http://arxiv.org/abs/nucl-ex/0410003} {arXiv:nucl-ex/0410003 [nucl-ex]}
  \BibitemShut {NoStop}%
\bibitem [{\citenamefont {Back}\ \emph {et~al.}(2005)\citenamefont {Back} \emph
  {et~al.}}]{Back:2004je}%
  \BibitemOpen
  \bibfield  {author} {\bibinfo {author} {\bibfnamefont {B.~B.}\ \bibnamefont
  {Back}} \emph {et~al.},\ }\href {\doibase 10.1016/j.nuclphysa.2005.03.084}
  {\bibfield  {journal} {\bibinfo  {journal} {Nucl. Phys.}\ }\textbf {\bibinfo
  {volume} {A757}},\ \bibinfo {pages} {28} (\bibinfo {year} {2005})},\ \Eprint
  {http://arxiv.org/abs/nucl-ex/0410022} {arXiv:nucl-ex/0410022 [nucl-ex]}
  \BibitemShut {NoStop}%
\bibitem [{\citenamefont {Arsene}\ \emph {et~al.}(2005)\citenamefont {Arsene}
  \emph {et~al.}}]{Arsene:2004fa}%
  \BibitemOpen
  \bibfield  {author} {\bibinfo {author} {\bibfnamefont {I.}~\bibnamefont
  {Arsene}} \emph {et~al.} (\bibinfo {collaboration} {BRAHMS}),\ }\href
  {\doibase 10.1016/j.nuclphysa.2005.02.130} {\bibfield  {journal} {\bibinfo
  {journal} {Nucl. Phys.}\ }\textbf {\bibinfo {volume} {A757}},\ \bibinfo
  {pages} {1} (\bibinfo {year} {2005})},\ \Eprint
  {http://arxiv.org/abs/nucl-ex/0410020} {arXiv:nucl-ex/0410020 [nucl-ex]}
  \BibitemShut {NoStop}%
\bibitem [{\citenamefont {Heinz}\ and\ \citenamefont
  {Snellings}(2013)}]{Heinz:2013th}%
  \BibitemOpen
  \bibfield  {author} {\bibinfo {author} {\bibfnamefont {U.}~\bibnamefont
  {Heinz}}\ and\ \bibinfo {author} {\bibfnamefont {R.}~\bibnamefont
  {Snellings}},\ }\href {\doibase 10.1146/annurev-nucl-102212-170540}
  {\bibfield  {journal} {\bibinfo  {journal} {Ann. Rev. Nucl. Part. Sci.}\
  }\textbf {\bibinfo {volume} {63}},\ \bibinfo {pages} {123} (\bibinfo {year}
  {2013})},\ \Eprint {http://arxiv.org/abs/1301.2826} {arXiv:1301.2826
  [nucl-th]} \BibitemShut {NoStop}%
\bibitem [{\citenamefont {Teaney}(2010)}]{Teaney:2009qa}%
  \BibitemOpen
  \bibfield  {author} {\bibinfo {author} {\bibfnamefont {D.~A.}\ \bibnamefont
  {Teaney}},\ }in\ \href {\doibase 10.1142/9789814293297_0004} {\emph {\bibinfo
  {booktitle} {Quark-gluon plasma 4}}},\ \bibinfo {editor} {edited by\ \bibinfo
  {editor} {\bibfnamefont {R.~C.}\ \bibnamefont {Hwa}}\ and\ \bibinfo {editor}
  {\bibfnamefont {X.-N.}\ \bibnamefont {Wang}}}\ (\bibinfo {year} {2010})\ pp.\
  \bibinfo {pages} {207--266},\ \Eprint {http://arxiv.org/abs/0905.2433}
  {arXiv:0905.2433 [nucl-th]} \BibitemShut {NoStop}%
\bibitem [{\citenamefont {Luzum}\ and\ \citenamefont
  {Petersen}(2014)}]{Luzum:2013yya}%
  \BibitemOpen
  \bibfield  {author} {\bibinfo {author} {\bibfnamefont {M.}~\bibnamefont
  {Luzum}}\ and\ \bibinfo {author} {\bibfnamefont {H.}~\bibnamefont
  {Petersen}},\ }\href {\doibase 10.1088/0954-3899/41/6/063102} {\bibfield
  {journal} {\bibinfo  {journal} {J. Phys.}\ }\textbf {\bibinfo {volume}
  {G41}},\ \bibinfo {pages} {063102} (\bibinfo {year} {2014})},\ \Eprint
  {http://arxiv.org/abs/1312.5503} {arXiv:1312.5503 [nucl-th]} \BibitemShut
  {NoStop}%
\bibitem [{\citenamefont {Gale}\ \emph {et~al.}(2013)\citenamefont {Gale},
  \citenamefont {Jeon},\ and\ \citenamefont {Schenke}}]{Gale:2013da}%
  \BibitemOpen
  \bibfield  {author} {\bibinfo {author} {\bibfnamefont {C.}~\bibnamefont
  {Gale}}, \bibinfo {author} {\bibfnamefont {S.}~\bibnamefont {Jeon}}, \ and\
  \bibinfo {author} {\bibfnamefont {B.}~\bibnamefont {Schenke}},\ }\href
  {\doibase 10.1142/S0217751X13400113} {\bibfield  {journal} {\bibinfo
  {journal} {Int. J. Mod. Phys.}\ }\textbf {\bibinfo {volume} {A28}},\ \bibinfo
  {pages} {1340011} (\bibinfo {year} {2013})},\ \Eprint
  {http://arxiv.org/abs/1301.5893} {arXiv:1301.5893 [nucl-th]} \BibitemShut
  {NoStop}%
\bibitem [{\citenamefont {Derradi~de Souza}\ \emph {et~al.}(2016)\citenamefont
  {Derradi~de Souza}, \citenamefont {Koide},\ and\ \citenamefont
  {Kodama}}]{deSouza:2015ena}%
  \BibitemOpen
  \bibfield  {author} {\bibinfo {author} {\bibfnamefont {R.}~\bibnamefont
  {Derradi~de Souza}}, \bibinfo {author} {\bibfnamefont {T.}~\bibnamefont
  {Koide}}, \ and\ \bibinfo {author} {\bibfnamefont {T.}~\bibnamefont
  {Kodama}},\ }\href {\doibase 10.1016/j.ppnp.2015.09.002} {\bibfield
  {journal} {\bibinfo  {journal} {Prog. Part. Nucl. Phys.}\ }\textbf {\bibinfo
  {volume} {86}},\ \bibinfo {pages} {35} (\bibinfo {year} {2016})},\ \Eprint
  {http://arxiv.org/abs/1506.03863} {arXiv:1506.03863 [nucl-th]} \BibitemShut
  {NoStop}%
\bibitem [{\citenamefont {Arnold}\ \emph {et~al.}(2003)\citenamefont {Arnold},
  \citenamefont {Moore},\ and\ \citenamefont {Yaffe}}]{Arnold:2002zm}%
  \BibitemOpen
  \bibfield  {author} {\bibinfo {author} {\bibfnamefont {P.~B.}\ \bibnamefont
  {Arnold}}, \bibinfo {author} {\bibfnamefont {G.~D.}\ \bibnamefont {Moore}}, \
  and\ \bibinfo {author} {\bibfnamefont {L.~G.}\ \bibnamefont {Yaffe}},\ }\href
  {\doibase 10.1088/1126-6708/2003/01/030} {\bibfield  {journal} {\bibinfo
  {journal} {JHEP}\ }\textbf {\bibinfo {volume} {01}},\ \bibinfo {pages} {030}
  (\bibinfo {year} {2003})},\ \Eprint {http://arxiv.org/abs/hep-ph/0209353}
  {arXiv:hep-ph/0209353 [hep-ph]} \BibitemShut {NoStop}%
\bibitem [{\citenamefont {Iancu}\ \emph {et~al.}(2002)\citenamefont {Iancu},
  \citenamefont {Leonidov},\ and\ \citenamefont {McLerran}}]{Iancu:2002xk}%
  \BibitemOpen
  \bibfield  {author} {\bibinfo {author} {\bibfnamefont {E.}~\bibnamefont
  {Iancu}}, \bibinfo {author} {\bibfnamefont {A.}~\bibnamefont {Leonidov}}, \
  and\ \bibinfo {author} {\bibfnamefont {L.}~\bibnamefont {McLerran}},\ }in\
  \href@noop {} {\emph {\bibinfo {booktitle} {{QCD perspectives on hot and
  dense matter. Proceedings, NATO Advanced Study Institute, Summer School,
  Cargese, France, August 6-18, 2001}}}}\ (\bibinfo {year} {2002})\ pp.\
  \bibinfo {pages} {73--145},\ \Eprint {http://arxiv.org/abs/hep-ph/0202270}
  {arXiv:hep-ph/0202270 [hep-ph]} \BibitemShut {NoStop}%
\bibitem [{\citenamefont {Iancu}\ and\ \citenamefont
  {Venugopalan}(2003)}]{Iancu:2003xm}%
  \BibitemOpen
  \bibfield  {author} {\bibinfo {author} {\bibfnamefont {E.}~\bibnamefont
  {Iancu}}\ and\ \bibinfo {author} {\bibfnamefont {R.}~\bibnamefont
  {Venugopalan}},\ }in\ \href {\doibase 10.1142/9789812795533_0005} {\emph
  {\bibinfo {booktitle} {Quark-gluon plasma 4}}},\ \bibinfo {editor} {edited
  by\ \bibinfo {editor} {\bibfnamefont {R.~C.}\ \bibnamefont {Hwa}}\ and\
  \bibinfo {editor} {\bibfnamefont {X.-N.}\ \bibnamefont {Wang}}}\ (\bibinfo
  {year} {2003})\ pp.\ \bibinfo {pages} {249--3363},\ \Eprint
  {http://arxiv.org/abs/hep-ph/0303204} {arXiv:hep-ph/0303204 [hep-ph]}
  \BibitemShut {NoStop}%
\bibitem [{\citenamefont {Gelis}\ \emph {et~al.}(2010)\citenamefont {Gelis},
  \citenamefont {Iancu}, \citenamefont {Jalilian-Marian},\ and\ \citenamefont
  {Venugopalan}}]{Gelis:2010nm}%
  \BibitemOpen
  \bibfield  {author} {\bibinfo {author} {\bibfnamefont {F.}~\bibnamefont
  {Gelis}}, \bibinfo {author} {\bibfnamefont {E.}~\bibnamefont {Iancu}},
  \bibinfo {author} {\bibfnamefont {J.}~\bibnamefont {Jalilian-Marian}}, \ and\
  \bibinfo {author} {\bibfnamefont {R.}~\bibnamefont {Venugopalan}},\ }\href
  {\doibase 10.1146/annurev.nucl.010909.083629} {\bibfield  {journal} {\bibinfo
   {journal} {Ann. Rev. Nucl. Part. Sci.}\ }\textbf {\bibinfo {volume} {60}},\
  \bibinfo {pages} {463} (\bibinfo {year} {2010})},\ \Eprint
  {http://arxiv.org/abs/1002.0333} {arXiv:1002.0333 [hep-ph]} \BibitemShut
  {NoStop}%
\bibitem [{\citenamefont {Gelis}\ \emph {et~al.}(2007)\citenamefont {Gelis},
  \citenamefont {Lappi},\ and\ \citenamefont {Venugopalan}}]{Gelis:2007kn}%
  \BibitemOpen
  \bibfield  {author} {\bibinfo {author} {\bibfnamefont {F.}~\bibnamefont
  {Gelis}}, \bibinfo {author} {\bibfnamefont {T.}~\bibnamefont {Lappi}}, \ and\
  \bibinfo {author} {\bibfnamefont {R.}~\bibnamefont {Venugopalan}},\
  }\bibfield  {booktitle} {\emph {\bibinfo {booktitle} {{Hadron physics.
  Proceedings, 10th International Workshop, Florianopolis, Brazil, April 26-31,
  2007}}},\ }\href {\doibase 10.1142/S0218301307008331} {\bibfield  {journal}
  {\bibinfo  {journal} {Int. J. Mod. Phys.}\ }\textbf {\bibinfo {volume}
  {E16}},\ \bibinfo {pages} {2595} (\bibinfo {year} {2007})},\ \Eprint
  {http://arxiv.org/abs/0708.0047} {arXiv:0708.0047 [hep-ph]} \BibitemShut
  {NoStop}%
\bibitem [{\citenamefont {Berges}\ \emph
  {et~al.}(2014{\natexlab{a}})\citenamefont {Berges}, \citenamefont
  {Boguslavski}, \citenamefont {Schlichting},\ and\ \citenamefont
  {Venugopalan}}]{Berges:2013fga}%
  \BibitemOpen
  \bibfield  {author} {\bibinfo {author} {\bibfnamefont {J.}~\bibnamefont
  {Berges}}, \bibinfo {author} {\bibfnamefont {K.}~\bibnamefont {Boguslavski}},
  \bibinfo {author} {\bibfnamefont {S.}~\bibnamefont {Schlichting}}, \ and\
  \bibinfo {author} {\bibfnamefont {R.}~\bibnamefont {Venugopalan}},\ }\href
  {\doibase 10.1103/PhysRevD.89.114007} {\bibfield  {journal} {\bibinfo
  {journal} {Phys. Rev.}\ }\textbf {\bibinfo {volume} {D89}},\ \bibinfo {pages}
  {114007} (\bibinfo {year} {2014}{\natexlab{a}})},\ \Eprint
  {http://arxiv.org/abs/1311.3005} {arXiv:1311.3005 [hep-ph]} \BibitemShut
  {NoStop}%
\bibitem [{\citenamefont {Berges}\ \emph
  {et~al.}(2014{\natexlab{b}})\citenamefont {Berges}, \citenamefont {Schenke},
  \citenamefont {Schlichting},\ and\ \citenamefont
  {Venugopalan}}]{Berges:2014yta}%
  \BibitemOpen
  \bibfield  {author} {\bibinfo {author} {\bibfnamefont {J.}~\bibnamefont
  {Berges}}, \bibinfo {author} {\bibfnamefont {B.}~\bibnamefont {Schenke}},
  \bibinfo {author} {\bibfnamefont {S.}~\bibnamefont {Schlichting}}, \ and\
  \bibinfo {author} {\bibfnamefont {R.}~\bibnamefont {Venugopalan}},\
  }\bibfield  {booktitle} {\emph {\bibinfo {booktitle} {{Proceedings, 24th
  International Conference on Ultra-Relativistic Nucleus-Nucleus Collisions
  (Quark Matter 2014): Darmstadt, Germany, May 19-24, 2014}}},\ }\href
  {\doibase 10.1016/j.nuclphysa.2014.08.103} {\bibfield  {journal} {\bibinfo
  {journal} {Nucl. Phys.}\ }\textbf {\bibinfo {volume} {A931}},\ \bibinfo
  {pages} {348} (\bibinfo {year} {2014}{\natexlab{b}})},\ \Eprint
  {http://arxiv.org/abs/1409.1638} {arXiv:1409.1638 [hep-ph]} \BibitemShut
  {NoStop}%
\bibitem [{\citenamefont {Berges}\ \emph
  {et~al.}(2014{\natexlab{c}})\citenamefont {Berges}, \citenamefont
  {Boguslavski}, \citenamefont {Schlichting},\ and\ \citenamefont
  {Venugopalan}}]{Berges:2013eia}%
  \BibitemOpen
  \bibfield  {author} {\bibinfo {author} {\bibfnamefont {J.}~\bibnamefont
  {Berges}}, \bibinfo {author} {\bibfnamefont {K.}~\bibnamefont {Boguslavski}},
  \bibinfo {author} {\bibfnamefont {S.}~\bibnamefont {Schlichting}}, \ and\
  \bibinfo {author} {\bibfnamefont {R.}~\bibnamefont {Venugopalan}},\ }\href
  {\doibase 10.1103/PhysRevD.89.074011} {\bibfield  {journal} {\bibinfo
  {journal} {Phys. Rev.}\ }\textbf {\bibinfo {volume} {D89}},\ \bibinfo {pages}
  {074011} (\bibinfo {year} {2014}{\natexlab{c}})},\ \Eprint
  {http://arxiv.org/abs/1303.5650} {arXiv:1303.5650 [hep-ph]} \BibitemShut
  {NoStop}%
\bibitem [{\citenamefont {Baier}\ \emph {et~al.}(2001)\citenamefont {Baier},
  \citenamefont {Mueller}, \citenamefont {Schiff},\ and\ \citenamefont
  {Son}}]{Baier:2000sb}%
  \BibitemOpen
  \bibfield  {author} {\bibinfo {author} {\bibfnamefont {R.}~\bibnamefont
  {Baier}}, \bibinfo {author} {\bibfnamefont {A.~H.}\ \bibnamefont {Mueller}},
  \bibinfo {author} {\bibfnamefont {D.}~\bibnamefont {Schiff}}, \ and\ \bibinfo
  {author} {\bibfnamefont {D.~T.}\ \bibnamefont {Son}},\ }\href {\doibase
  10.1016/S0370-2693(01)00191-5} {\bibfield  {journal} {\bibinfo  {journal}
  {Phys. Lett.}\ }\textbf {\bibinfo {volume} {B502}},\ \bibinfo {pages} {51}
  (\bibinfo {year} {2001})},\ \Eprint {http://arxiv.org/abs/hep-ph/0009237}
  {arXiv:hep-ph/0009237 [hep-ph]} \BibitemShut {NoStop}%
\bibitem [{\citenamefont {Krasnitz}\ \emph {et~al.}(2003)\citenamefont
  {Krasnitz}, \citenamefont {Nara},\ and\ \citenamefont
  {Venugopalan}}]{Krasnitz:2003jw}%
  \BibitemOpen
  \bibfield  {author} {\bibinfo {author} {\bibfnamefont {A.}~\bibnamefont
  {Krasnitz}}, \bibinfo {author} {\bibfnamefont {Y.}~\bibnamefont {Nara}}, \
  and\ \bibinfo {author} {\bibfnamefont {R.}~\bibnamefont {Venugopalan}},\
  }\href {\doibase 10.1016/j.nuclphysa.2003.08.004} {\bibfield  {journal}
  {\bibinfo  {journal} {Nucl. Phys.}\ }\textbf {\bibinfo {volume} {A727}},\
  \bibinfo {pages} {427} (\bibinfo {year} {2003})},\ \Eprint
  {http://arxiv.org/abs/hep-ph/0305112} {arXiv:hep-ph/0305112 [hep-ph]}
  \BibitemShut {NoStop}%
\bibitem [{\citenamefont {Lappi}\ and\ \citenamefont
  {McLerran}(2006)}]{Lappi:2006fp}%
  \BibitemOpen
  \bibfield  {author} {\bibinfo {author} {\bibfnamefont {T.}~\bibnamefont
  {Lappi}}\ and\ \bibinfo {author} {\bibfnamefont {L.}~\bibnamefont
  {McLerran}},\ }\href {\doibase 10.1016/j.nuclphysa.2006.04.001} {\bibfield
  {journal} {\bibinfo  {journal} {Nucl. Phys.}\ }\textbf {\bibinfo {volume}
  {A772}},\ \bibinfo {pages} {200} (\bibinfo {year} {2006})},\ \Eprint
  {http://arxiv.org/abs/hep-ph/0602189} {arXiv:hep-ph/0602189 [hep-ph]}
  \BibitemShut {NoStop}%
\bibitem [{\citenamefont {Lappi}(2011)}]{Lappi:2011ju}%
  \BibitemOpen
  \bibfield  {author} {\bibinfo {author} {\bibfnamefont {T.}~\bibnamefont
  {Lappi}},\ }\href {\doibase 10.1016/j.physletb.2011.08.011} {\bibfield
  {journal} {\bibinfo  {journal} {Phys. Lett.}\ }\textbf {\bibinfo {volume}
  {B703}},\ \bibinfo {pages} {325} (\bibinfo {year} {2011})},\ \Eprint
  {http://arxiv.org/abs/1105.5511} {arXiv:1105.5511 [hep-ph]} \BibitemShut
  {NoStop}%
\bibitem [{\citenamefont {Epelbaum}\ and\ \citenamefont
  {Gelis}(2013{\natexlab{a}})}]{Epelbaum:2013waa}%
  \BibitemOpen
  \bibfield  {author} {\bibinfo {author} {\bibfnamefont {T.}~\bibnamefont
  {Epelbaum}}\ and\ \bibinfo {author} {\bibfnamefont {F.}~\bibnamefont
  {Gelis}},\ }\href {\doibase 10.1103/PhysRevD.88.085015} {\bibfield  {journal}
  {\bibinfo  {journal} {Phys. Rev.}\ }\textbf {\bibinfo {volume} {D88}},\
  \bibinfo {pages} {085015} (\bibinfo {year} {2013}{\natexlab{a}})},\ \Eprint
  {http://arxiv.org/abs/1307.1765} {arXiv:1307.1765 [hep-ph]} \BibitemShut
  {NoStop}%
\bibitem [{\citenamefont {Epelbaum}\ and\ \citenamefont
  {Gelis}(2013{\natexlab{b}})}]{Gelis:2013rba}%
  \BibitemOpen
  \bibfield  {author} {\bibinfo {author} {\bibfnamefont {T.}~\bibnamefont
  {Epelbaum}}\ and\ \bibinfo {author} {\bibfnamefont {F.}~\bibnamefont
  {Gelis}},\ }\href {\doibase 10.1103/PhysRevLett.111.232301} {\bibfield
  {journal} {\bibinfo  {journal} {Phys. Rev. Lett.}\ }\textbf {\bibinfo
  {volume} {111}},\ \bibinfo {pages} {232301} (\bibinfo {year}
  {2013}{\natexlab{b}})},\ \Eprint {http://arxiv.org/abs/1307.2214}
  {arXiv:1307.2214 [hep-ph]} \BibitemShut {NoStop}%
\bibitem [{\citenamefont {Schenke}\ \emph {et~al.}(2015)\citenamefont
  {Schenke}, \citenamefont {Schlichting},\ and\ \citenamefont
  {Venugopalan}}]{Schenke:2015aqa}%
  \BibitemOpen
  \bibfield  {author} {\bibinfo {author} {\bibfnamefont {B.}~\bibnamefont
  {Schenke}}, \bibinfo {author} {\bibfnamefont {S.}~\bibnamefont
  {Schlichting}}, \ and\ \bibinfo {author} {\bibfnamefont {R.}~\bibnamefont
  {Venugopalan}},\ }\href {\doibase 10.1016/j.physletb.2015.05.051} {\bibfield
  {journal} {\bibinfo  {journal} {Phys. Lett.}\ }\textbf {\bibinfo {volume}
  {B747}},\ \bibinfo {pages} {76} (\bibinfo {year} {2015})},\ \Eprint
  {http://arxiv.org/abs/1502.01331} {arXiv:1502.01331 [hep-ph]} \BibitemShut
  {NoStop}%
\bibitem [{\citenamefont {Xu}\ and\ \citenamefont {Greiner}(2005)}]{Xu:2004mz}%
  \BibitemOpen
  \bibfield  {author} {\bibinfo {author} {\bibfnamefont {Z.}~\bibnamefont
  {Xu}}\ and\ \bibinfo {author} {\bibfnamefont {C.}~\bibnamefont {Greiner}},\
  }\href {\doibase 10.1103/PhysRevC.71.064901} {\bibfield  {journal} {\bibinfo
  {journal} {Phys. Rev.}\ }\textbf {\bibinfo {volume} {C71}},\ \bibinfo {pages}
  {064901} (\bibinfo {year} {2005})},\ \Eprint
  {http://arxiv.org/abs/hep-ph/0406278} {arXiv:hep-ph/0406278 [hep-ph]}
  \BibitemShut {NoStop}%
\bibitem [{\citenamefont {El}\ \emph {et~al.}(2008)\citenamefont {El},
  \citenamefont {Xu},\ and\ \citenamefont {Greiner}}]{El:2007vg}%
  \BibitemOpen
  \bibfield  {author} {\bibinfo {author} {\bibfnamefont {A.}~\bibnamefont
  {El}}, \bibinfo {author} {\bibfnamefont {Z.}~\bibnamefont {Xu}}, \ and\
  \bibinfo {author} {\bibfnamefont {C.}~\bibnamefont {Greiner}},\ }\href
  {\doibase 10.1016/j.nuclphysa.2008.03.005} {\bibfield  {journal} {\bibinfo
  {journal} {Nucl. Phys.}\ }\textbf {\bibinfo {volume} {A806}},\ \bibinfo
  {pages} {287} (\bibinfo {year} {2008})},\ \Eprint
  {http://arxiv.org/abs/0712.3734} {arXiv:0712.3734 [hep-ph]} \BibitemShut
  {NoStop}%
\bibitem [{\citenamefont {Greif}\ \emph {et~al.}(2017)\citenamefont {Greif},
  \citenamefont {Greiner}, \citenamefont {Schenke}, \citenamefont
  {Schlichting},\ and\ \citenamefont {Xu}}]{Greif:2017bnr}%
  \BibitemOpen
  \bibfield  {author} {\bibinfo {author} {\bibfnamefont {M.}~\bibnamefont
  {Greif}}, \bibinfo {author} {\bibfnamefont {C.}~\bibnamefont {Greiner}},
  \bibinfo {author} {\bibfnamefont {B.}~\bibnamefont {Schenke}}, \bibinfo
  {author} {\bibfnamefont {S.}~\bibnamefont {Schlichting}}, \ and\ \bibinfo
  {author} {\bibfnamefont {Z.}~\bibnamefont {Xu}},\ }\href {\doibase
  10.1103/PhysRevD.96.091504} {\bibfield  {journal} {\bibinfo  {journal} {Phys.
  Rev.}\ }\textbf {\bibinfo {volume} {D96}},\ \bibinfo {pages} {091504}
  (\bibinfo {year} {2017})},\ \Eprint {http://arxiv.org/abs/1708.02076}
  {arXiv:1708.02076 [hep-ph]} \BibitemShut {NoStop}%
\bibitem [{\citenamefont {Abraao~York}\ \emph {et~al.}(2014)\citenamefont
  {Abraao~York}, \citenamefont {Kurkela}, \citenamefont {Lu},\ and\
  \citenamefont {Moore}}]{York:2014wja}%
  \BibitemOpen
  \bibfield  {author} {\bibinfo {author} {\bibfnamefont {M.~C.}\ \bibnamefont
  {Abraao~York}}, \bibinfo {author} {\bibfnamefont {A.}~\bibnamefont
  {Kurkela}}, \bibinfo {author} {\bibfnamefont {E.}~\bibnamefont {Lu}}, \ and\
  \bibinfo {author} {\bibfnamefont {G.~D.}\ \bibnamefont {Moore}},\ }\href
  {\doibase 10.1103/PhysRevD.89.074036} {\bibfield  {journal} {\bibinfo
  {journal} {Phys. Rev.}\ }\textbf {\bibinfo {volume} {D89}},\ \bibinfo {pages}
  {074036} (\bibinfo {year} {2014})},\ \Eprint {http://arxiv.org/abs/1401.3751}
  {arXiv:1401.3751 [hep-ph]} \BibitemShut {NoStop}%
\bibitem [{\citenamefont {Kurkela}\ and\ \citenamefont
  {Lu}(2014)}]{Kurkela:2014tea}%
  \BibitemOpen
  \bibfield  {author} {\bibinfo {author} {\bibfnamefont {A.}~\bibnamefont
  {Kurkela}}\ and\ \bibinfo {author} {\bibfnamefont {E.}~\bibnamefont {Lu}},\
  }\href {\doibase 10.1103/PhysRevLett.113.182301} {\bibfield  {journal}
  {\bibinfo  {journal} {Phys. Rev. Lett.}\ }\textbf {\bibinfo {volume} {113}},\
  \bibinfo {pages} {182301} (\bibinfo {year} {2014})},\ \Eprint
  {http://arxiv.org/abs/1405.6318} {arXiv:1405.6318 [hep-ph]} \BibitemShut
  {NoStop}%
\bibitem [{\citenamefont {Kurkela}\ and\ \citenamefont
  {Zhu}(2015)}]{Kurkela:2015qoa}%
  \BibitemOpen
  \bibfield  {author} {\bibinfo {author} {\bibfnamefont {A.}~\bibnamefont
  {Kurkela}}\ and\ \bibinfo {author} {\bibfnamefont {Y.}~\bibnamefont {Zhu}},\
  }\href {\doibase 10.1103/PhysRevLett.115.182301} {\bibfield  {journal}
  {\bibinfo  {journal} {Phys. Rev. Lett.}\ }\textbf {\bibinfo {volume} {115}},\
  \bibinfo {pages} {182301} (\bibinfo {year} {2015})},\ \Eprint
  {http://arxiv.org/abs/1506.06647} {arXiv:1506.06647 [hep-ph]} \BibitemShut
  {NoStop}%
\bibitem [{\citenamefont {Keegan}\ \emph {et~al.}(2016)\citenamefont {Keegan},
  \citenamefont {Kurkela}, \citenamefont {Mazeliauskas},\ and\ \citenamefont
  {Teaney}}]{Keegan:2016cpi}%
  \BibitemOpen
  \bibfield  {author} {\bibinfo {author} {\bibfnamefont {L.}~\bibnamefont
  {Keegan}}, \bibinfo {author} {\bibfnamefont {A.}~\bibnamefont {Kurkela}},
  \bibinfo {author} {\bibfnamefont {A.}~\bibnamefont {Mazeliauskas}}, \ and\
  \bibinfo {author} {\bibfnamefont {D.}~\bibnamefont {Teaney}},\ }\href
  {\doibase 10.1007/JHEP08(2016)171} {\bibfield  {journal} {\bibinfo  {journal}
  {JHEP}\ }\textbf {\bibinfo {volume} {08}},\ \bibinfo {pages} {171} (\bibinfo
  {year} {2016})},\ \Eprint {http://arxiv.org/abs/1605.04287} {arXiv:1605.04287
  [hep-ph]} \BibitemShut {NoStop}%
\bibitem [{\citenamefont {Baier}\ \emph {et~al.}(1997)\citenamefont {Baier},
  \citenamefont {Dokshitzer}, \citenamefont {Mueller}, \citenamefont {Peigne},\
  and\ \citenamefont {Schiff}}]{Baier:1996kr}%
  \BibitemOpen
  \bibfield  {author} {\bibinfo {author} {\bibfnamefont {R.}~\bibnamefont
  {Baier}}, \bibinfo {author} {\bibfnamefont {Y.~L.}\ \bibnamefont
  {Dokshitzer}}, \bibinfo {author} {\bibfnamefont {A.~H.}\ \bibnamefont
  {Mueller}}, \bibinfo {author} {\bibfnamefont {S.}~\bibnamefont {Peigne}}, \
  and\ \bibinfo {author} {\bibfnamefont {D.}~\bibnamefont {Schiff}},\ }\href
  {\doibase 10.1016/S0550-3213(96)00553-6} {\bibfield  {journal} {\bibinfo
  {journal} {Nucl. Phys.}\ }\textbf {\bibinfo {volume} {B483}},\ \bibinfo
  {pages} {291} (\bibinfo {year} {1997})},\ \Eprint
  {http://arxiv.org/abs/hep-ph/9607355} {arXiv:hep-ph/9607355 [hep-ph]}
  \BibitemShut {NoStop}%
\bibitem [{\citenamefont {Kurkela}\ \emph {et~al.}(2019)\citenamefont
  {Kurkela}, \citenamefont {Mazeliauskas}, \citenamefont {Paquet},
  \citenamefont {Schlichting},\ and\ \citenamefont {Teaney}}]{Kurkela:2018vqr}%
  \BibitemOpen
  \bibfield  {author} {\bibinfo {author} {\bibfnamefont {A.}~\bibnamefont
  {Kurkela}}, \bibinfo {author} {\bibfnamefont {A.}~\bibnamefont
  {Mazeliauskas}}, \bibinfo {author} {\bibfnamefont {J.-F.}\ \bibnamefont
  {Paquet}}, \bibinfo {author} {\bibfnamefont {S.}~\bibnamefont {Schlichting}},
  \ and\ \bibinfo {author} {\bibfnamefont {D.}~\bibnamefont {Teaney}},\ }\href
  {\doibase 10.1103/PhysRevC.99.034910} {\bibfield  {journal} {\bibinfo
  {journal} {Phys. Rev.}\ }\textbf {\bibinfo {volume} {C99}},\ \bibinfo {pages}
  {034910} (\bibinfo {year} {2019})},\ \Eprint
  {http://arxiv.org/abs/1805.00961} {arXiv:1805.00961 [hep-ph]} \BibitemShut
  {NoStop}%
\bibitem [{\citenamefont {Kurkela}\ \emph {et~al.}(2018)\citenamefont
  {Kurkela}, \citenamefont {Mazeliauskas}, \citenamefont {Paquet},
  \citenamefont {Schlichting},\ and\ \citenamefont {Teaney}}]{kompost_github}%
  \BibitemOpen
  \bibfield  {author} {\bibinfo {author} {\bibfnamefont {A.}~\bibnamefont
  {Kurkela}}, \bibinfo {author} {\bibfnamefont {A.}~\bibnamefont
  {Mazeliauskas}}, \bibinfo {author} {\bibfnamefont {J.-F.}\ \bibnamefont
  {Paquet}}, \bibinfo {author} {\bibfnamefont {S.}~\bibnamefont {Schlichting}},
  \ and\ \bibinfo {author} {\bibfnamefont {D.}~\bibnamefont {Teaney}},\ }\href
  {https://github.com/KMPST/KoMPoST} {\enquote {\bibinfo {title}
  {{K{\o}MP{\o}ST}: linearized kinetic theory propagator of initial conditions
  for heavy ion collisions},}\ }\bibinfo {howpublished}
  {https://github.com/KMPST/KoMPoST} (\bibinfo {year} {2018})\BibitemShut
  {NoStop}%
\bibitem [{Note1()}]{Note1}%
  \BibitemOpen
  \bibinfo {note} {Note that in small collision systems such as $p$+$p$ or
  $p$+Pb, a separation of scales is not warranted, and the linearized
  description underlying K{\o }MP{\o }ST{} may be inapplicable}\BibitemShut
  {NoStop}%
\bibitem [{\citenamefont {Heller}\ \emph {et~al.}(2018)\citenamefont {Heller},
  \citenamefont {Kurkela}, \citenamefont {Spaliński},\ and\ \citenamefont
  {Svensson}}]{Heller:2016rtz}%
  \BibitemOpen
  \bibfield  {author} {\bibinfo {author} {\bibfnamefont {M.~P.}\ \bibnamefont
  {Heller}}, \bibinfo {author} {\bibfnamefont {A.}~\bibnamefont {Kurkela}},
  \bibinfo {author} {\bibfnamefont {M.}~\bibnamefont {Spaliński}}, \ and\
  \bibinfo {author} {\bibfnamefont {V.}~\bibnamefont {Svensson}},\ }\href
  {\doibase 10.1103/PhysRevD.97.091503} {\bibfield  {journal} {\bibinfo
  {journal} {Phys. Rev.}\ }\textbf {\bibinfo {volume} {D97}},\ \bibinfo {pages}
  {091503} (\bibinfo {year} {2018})},\ \Eprint
  {http://arxiv.org/abs/1609.04803} {arXiv:1609.04803 [nucl-th]} \BibitemShut
  {NoStop}%
\bibitem [{\citenamefont {Romatschke}(2018)}]{Romatschke:2017vte}%
  \BibitemOpen
  \bibfield  {author} {\bibinfo {author} {\bibfnamefont {P.}~\bibnamefont
  {Romatschke}},\ }\href {\doibase 10.1103/PhysRevLett.120.012301} {\bibfield
  {journal} {\bibinfo  {journal} {Phys. Rev. Lett.}\ }\textbf {\bibinfo
  {volume} {120}},\ \bibinfo {pages} {012301} (\bibinfo {year} {2018})},\
  \Eprint {http://arxiv.org/abs/1704.08699} {arXiv:1704.08699 [hep-th]}
  \BibitemShut {NoStop}%
\bibitem [{\citenamefont {Strickland}\ \emph {et~al.}(2018)\citenamefont
  {Strickland}, \citenamefont {Noronha},\ and\ \citenamefont
  {Denicol}}]{Strickland:2017kux}%
  \BibitemOpen
  \bibfield  {author} {\bibinfo {author} {\bibfnamefont {M.}~\bibnamefont
  {Strickland}}, \bibinfo {author} {\bibfnamefont {J.}~\bibnamefont {Noronha}},
  \ and\ \bibinfo {author} {\bibfnamefont {G.}~\bibnamefont {Denicol}},\ }\href
  {\doibase 10.1103/PhysRevD.97.036020} {\bibfield  {journal} {\bibinfo
  {journal} {Phys. Rev.}\ }\textbf {\bibinfo {volume} {D97}},\ \bibinfo {pages}
  {036020} (\bibinfo {year} {2018})},\ \Eprint
  {http://arxiv.org/abs/1709.06644} {arXiv:1709.06644 [nucl-th]} \BibitemShut
  {NoStop}%
\bibitem [{\citenamefont {Behtash}\ \emph {et~al.}(2018)\citenamefont
  {Behtash}, \citenamefont {Cruz-Camacho},\ and\ \citenamefont
  {Martinez}}]{Behtash:2017wqg}%
  \BibitemOpen
  \bibfield  {author} {\bibinfo {author} {\bibfnamefont {A.}~\bibnamefont
  {Behtash}}, \bibinfo {author} {\bibfnamefont {C.~N.}\ \bibnamefont
  {Cruz-Camacho}}, \ and\ \bibinfo {author} {\bibfnamefont {M.}~\bibnamefont
  {Martinez}},\ }\href {\doibase 10.1103/PhysRevD.97.044041} {\bibfield
  {journal} {\bibinfo  {journal} {Phys. Rev.}\ }\textbf {\bibinfo {volume}
  {D97}},\ \bibinfo {pages} {044041} (\bibinfo {year} {2018})},\ \Eprint
  {http://arxiv.org/abs/1711.01745} {arXiv:1711.01745 [hep-th]} \BibitemShut
  {NoStop}%
\bibitem [{\citenamefont {Bazavov}\ \emph {et~al.}(2014)\citenamefont {Bazavov}
  \emph {et~al.}}]{Bazavov:2014pvz}%
  \BibitemOpen
  \bibfield  {author} {\bibinfo {author} {\bibfnamefont {A.}~\bibnamefont
  {Bazavov}} \emph {et~al.} (\bibinfo {collaboration} {HotQCD}),\ }\href
  {\doibase 10.1103/PhysRevD.90.094503} {\bibfield  {journal} {\bibinfo
  {journal} {Phys. Rev.}\ }\textbf {\bibinfo {volume} {D90}},\ \bibinfo {pages}
  {094503} (\bibinfo {year} {2014})},\ \Eprint {http://arxiv.org/abs/1407.6387}
  {arXiv:1407.6387 [hep-lat]} \BibitemShut {NoStop}%
\bibitem [{\citenamefont {Borsanyi}\ \emph {et~al.}(2016)\citenamefont
  {Borsanyi} \emph {et~al.}}]{Borsanyi:2016ksw}%
  \BibitemOpen
  \bibfield  {author} {\bibinfo {author} {\bibfnamefont {S.}~\bibnamefont
  {Borsanyi}} \emph {et~al.},\ }\href {\doibase 10.1038/nature20115} {\bibfield
   {journal} {\bibinfo  {journal} {Nature}\ }\textbf {\bibinfo {volume}
  {539}},\ \bibinfo {pages} {69} (\bibinfo {year} {2016})},\ \Eprint
  {http://arxiv.org/abs/1606.07494} {arXiv:1606.07494 [hep-lat]} \BibitemShut
  {NoStop}%
\bibitem [{Note2()}]{Note2}%
  \BibitemOpen
  \bibinfo {note} {Small scale fluctuations ---less than $\tau _\protect
  \textsc {ekt}$---were smeared from the IP-glasma initial conditions to limit
  the strain on the linearized perturbation approximation.}\BibitemShut {Stop}%
\bibitem [{\citenamefont {Schenke}\ \emph
  {et~al.}(2012{\natexlab{a}})\citenamefont {Schenke}, \citenamefont
  {Tribedy},\ and\ \citenamefont {Venugopalan}}]{Schenke:2012wb}%
  \BibitemOpen
  \bibfield  {author} {\bibinfo {author} {\bibfnamefont {B.}~\bibnamefont
  {Schenke}}, \bibinfo {author} {\bibfnamefont {P.}~\bibnamefont {Tribedy}}, \
  and\ \bibinfo {author} {\bibfnamefont {R.}~\bibnamefont {Venugopalan}},\
  }\href {\doibase 10.1103/PhysRevLett.108.252301} {\bibfield  {journal}
  {\bibinfo  {journal} {Phys. Rev. Lett.}\ }\textbf {\bibinfo {volume} {108}},\
  \bibinfo {pages} {252301} (\bibinfo {year} {2012}{\natexlab{a}})},\ \Eprint
  {http://arxiv.org/abs/1202.6646} {arXiv:1202.6646 [nucl-th]} \BibitemShut
  {NoStop}%
\bibitem [{\citenamefont {Schenke}\ \emph
  {et~al.}(2012{\natexlab{b}})\citenamefont {Schenke}, \citenamefont
  {Tribedy},\ and\ \citenamefont {Venugopalan}}]{Schenke:2012fw}%
  \BibitemOpen
  \bibfield  {author} {\bibinfo {author} {\bibfnamefont {B.}~\bibnamefont
  {Schenke}}, \bibinfo {author} {\bibfnamefont {P.}~\bibnamefont {Tribedy}}, \
  and\ \bibinfo {author} {\bibfnamefont {R.}~\bibnamefont {Venugopalan}},\
  }\href {\doibase 10.1103/PhysRevC.86.034908} {\bibfield  {journal} {\bibinfo
  {journal} {Phys. Rev.}\ }\textbf {\bibinfo {volume} {C86}},\ \bibinfo {pages}
  {034908} (\bibinfo {year} {2012}{\natexlab{b}})},\ \Eprint
  {http://arxiv.org/abs/1206.6805} {arXiv:1206.6805 [hep-ph]} \BibitemShut
  {NoStop}%
\bibitem [{Note3()}]{Note3}%
  \BibitemOpen
  \bibinfo {note} {We neglect perturbations in the $\delta T^{ij}$, $\delta
  T^{\eta i}$, and $\delta T^{\eta \eta }$ components, which are typically
  small and get further washed out during the nonequilibrium evolution, as they
  are not related to conserved quantities.}\BibitemShut {Stop}%
\bibitem [{\citenamefont {Schenke}\ \emph {et~al.}(2010)\citenamefont
  {Schenke}, \citenamefont {Jeon},\ and\ \citenamefont
  {Gale}}]{Schenke:2010nt}%
  \BibitemOpen
  \bibfield  {author} {\bibinfo {author} {\bibfnamefont {B.}~\bibnamefont
  {Schenke}}, \bibinfo {author} {\bibfnamefont {S.}~\bibnamefont {Jeon}}, \
  and\ \bibinfo {author} {\bibfnamefont {C.}~\bibnamefont {Gale}},\ }\href
  {\doibase 10.1103/PhysRevC.82.014903} {\bibfield  {journal} {\bibinfo
  {journal} {Phys. Rev.}\ }\textbf {\bibinfo {volume} {C82}},\ \bibinfo {pages}
  {014903} (\bibinfo {year} {2010})},\ \Eprint {http://arxiv.org/abs/1004.1408}
  {arXiv:1004.1408 [hep-ph]} \BibitemShut {NoStop}%
\bibitem [{\citenamefont {Schenke}\ \emph {et~al.}(2011)\citenamefont
  {Schenke}, \citenamefont {Jeon},\ and\ \citenamefont
  {Gale}}]{Schenke:2010rr}%
  \BibitemOpen
  \bibfield  {author} {\bibinfo {author} {\bibfnamefont {B.}~\bibnamefont
  {Schenke}}, \bibinfo {author} {\bibfnamefont {S.}~\bibnamefont {Jeon}}, \
  and\ \bibinfo {author} {\bibfnamefont {C.}~\bibnamefont {Gale}},\ }\href
  {\doibase 10.1103/PhysRevLett.106.042301} {\bibfield  {journal} {\bibinfo
  {journal} {Phys. Rev. Lett.}\ }\textbf {\bibinfo {volume} {106}},\ \bibinfo
  {pages} {042301} (\bibinfo {year} {2011})},\ \Eprint
  {http://arxiv.org/abs/1009.3244} {arXiv:1009.3244 [hep-ph]} \BibitemShut
  {NoStop}%
\bibitem [{\citenamefont {Paquet}\ \emph {et~al.}(2016)\citenamefont {Paquet},
  \citenamefont {Shen}, \citenamefont {Denicol}, \citenamefont {Luzum},
  \citenamefont {Schenke}, \citenamefont {Jeon},\ and\ \citenamefont
  {Gale}}]{Paquet:2015lta}%
  \BibitemOpen
  \bibfield  {author} {\bibinfo {author} {\bibfnamefont {J.-F.}\ \bibnamefont
  {Paquet}}, \bibinfo {author} {\bibfnamefont {C.}~\bibnamefont {Shen}},
  \bibinfo {author} {\bibfnamefont {G.~S.}\ \bibnamefont {Denicol}}, \bibinfo
  {author} {\bibfnamefont {M.}~\bibnamefont {Luzum}}, \bibinfo {author}
  {\bibfnamefont {B.}~\bibnamefont {Schenke}}, \bibinfo {author} {\bibfnamefont
  {S.}~\bibnamefont {Jeon}}, \ and\ \bibinfo {author} {\bibfnamefont
  {C.}~\bibnamefont {Gale}},\ }\href {\doibase 10.1103/PhysRevC.93.044906}
  {\bibfield  {journal} {\bibinfo  {journal} {Phys. Rev.}\ }\textbf {\bibinfo
  {volume} {C93}},\ \bibinfo {pages} {044906} (\bibinfo {year} {2016})},\
  \Eprint {http://arxiv.org/abs/1509.06738} {arXiv:1509.06738 [hep-ph]}
  \BibitemShut {NoStop}%
\bibitem [{\citenamefont {Huovinen}\ and\ \citenamefont
  {Petreczky}(2010)}]{Huovinen:2009yb}%
  \BibitemOpen
  \bibfield  {author} {\bibinfo {author} {\bibfnamefont {P.}~\bibnamefont
  {Huovinen}}\ and\ \bibinfo {author} {\bibfnamefont {P.}~\bibnamefont
  {Petreczky}},\ }\href {\doibase 10.1016/j.nuclphysa.2010.02.015} {\bibfield
  {journal} {\bibinfo  {journal} {Nucl. Phys.}\ }\textbf {\bibinfo {volume}
  {A837}},\ \bibinfo {pages} {26} (\bibinfo {year} {2010})},\ \Eprint
  {http://arxiv.org/abs/0912.2541} {arXiv:0912.2541 [hep-ph]} \BibitemShut
  {NoStop}%
\bibitem [{\citenamefont {Jackson}(1998)}]{Jackson:1998nia}%
  \BibitemOpen
  \bibfield  {author} {\bibinfo {author} {\bibfnamefont {J.~D.}\ \bibnamefont
  {Jackson}},\ }\href@noop {} {\emph {\bibinfo {title} {{Classical
  Electrodynamics}}}}\ (\bibinfo  {publisher} {Wiley},\ \bibinfo {year}
  {1998})\BibitemShut {NoStop}%
\bibitem [{\citenamefont {Vredevoogd}\ and\ \citenamefont
  {Pratt}(2009)}]{Vredevoogd:2008id}%
  \BibitemOpen
  \bibfield  {author} {\bibinfo {author} {\bibfnamefont {J.}~\bibnamefont
  {Vredevoogd}}\ and\ \bibinfo {author} {\bibfnamefont {S.}~\bibnamefont
  {Pratt}},\ }\href {\doibase 10.1103/PhysRevC.79.044915} {\bibfield  {journal}
  {\bibinfo  {journal} {Phys. Rev.}\ }\textbf {\bibinfo {volume} {C79}},\
  \bibinfo {pages} {044915} (\bibinfo {year} {2009})},\ \Eprint
  {http://arxiv.org/abs/0810.4325} {arXiv:0810.4325 [nucl-th]} \BibitemShut
  {NoStop}%
\bibitem [{\citenamefont {van~der Schee}\ \emph {et~al.}(2013)\citenamefont
  {van~der Schee}, \citenamefont {Romatschke},\ and\ \citenamefont
  {Pratt}}]{vanderSchee:2013pia}%
  \BibitemOpen
  \bibfield  {author} {\bibinfo {author} {\bibfnamefont {W.}~\bibnamefont
  {van~der Schee}}, \bibinfo {author} {\bibfnamefont {P.}~\bibnamefont
  {Romatschke}}, \ and\ \bibinfo {author} {\bibfnamefont {S.}~\bibnamefont
  {Pratt}},\ }\href {\doibase 10.1103/PhysRevLett.111.222302} {\bibfield
  {journal} {\bibinfo  {journal} {Phys. Rev. Lett.}\ }\textbf {\bibinfo
  {volume} {111}},\ \bibinfo {pages} {222302} (\bibinfo {year} {2013})},\
  \Eprint {http://arxiv.org/abs/1307.2539} {arXiv:1307.2539 [nucl-th]}
  \BibitemShut {NoStop}%
\bibitem [{\citenamefont {Romatschke}(2015)}]{Romatschke:2015gxa}%
  \BibitemOpen
  \bibfield  {author} {\bibinfo {author} {\bibfnamefont {P.}~\bibnamefont
  {Romatschke}},\ }\href {\doibase 10.1140/epjc/s10052-015-3509-3} {\bibfield
  {journal} {\bibinfo  {journal} {Eur. Phys. J.}\ }\textbf {\bibinfo {volume}
  {C75}},\ \bibinfo {pages} {305} (\bibinfo {year} {2015})},\ \Eprint
  {http://arxiv.org/abs/1502.04745} {arXiv:1502.04745 [nucl-th]} \BibitemShut
  {NoStop}%
\bibitem [{\citenamefont {Broniowski}\ \emph {et~al.}(2009)\citenamefont
  {Broniowski}, \citenamefont {Florkowski}, \citenamefont {Chojnacki},\ and\
  \citenamefont {Kisiel}}]{Broniowski:2008qk}%
  \BibitemOpen
  \bibfield  {author} {\bibinfo {author} {\bibfnamefont {W.}~\bibnamefont
  {Broniowski}}, \bibinfo {author} {\bibfnamefont {W.}~\bibnamefont
  {Florkowski}}, \bibinfo {author} {\bibfnamefont {M.}~\bibnamefont
  {Chojnacki}}, \ and\ \bibinfo {author} {\bibfnamefont {A.}~\bibnamefont
  {Kisiel}},\ }\href {\doibase 10.1103/PhysRevC.80.034902} {\bibfield
  {journal} {\bibinfo  {journal} {Phys. Rev.}\ }\textbf {\bibinfo {volume}
  {C80}},\ \bibinfo {pages} {034902} (\bibinfo {year} {2009})},\ \Eprint
  {http://arxiv.org/abs/0812.3393} {arXiv:0812.3393 [nucl-th]} \BibitemShut
  {NoStop}%
\bibitem [{\citenamefont {Liu}\ \emph {et~al.}(2015)\citenamefont {Liu},
  \citenamefont {Shen},\ and\ \citenamefont {Heinz}}]{Liu:2015nwa}%
  \BibitemOpen
  \bibfield  {author} {\bibinfo {author} {\bibfnamefont {J.}~\bibnamefont
  {Liu}}, \bibinfo {author} {\bibfnamefont {C.}~\bibnamefont {Shen}}, \ and\
  \bibinfo {author} {\bibfnamefont {U.}~\bibnamefont {Heinz}},\ }\href
  {\doibase 10.1103/PhysRevC.92.049904, 10.1103/PhysRevC.91.064906} {\bibfield
  {journal} {\bibinfo  {journal} {Phys. Rev.}\ }\textbf {\bibinfo {volume}
  {C91}},\ \bibinfo {pages} {064906} (\bibinfo {year} {2015})},\ \bibinfo
  {note} {[Erratum: Phys. Rev.C92,no.4,049904(2015)]},\ \Eprint
  {http://arxiv.org/abs/1504.02160} {arXiv:1504.02160 [nucl-th]} \BibitemShut
  {NoStop}%
\bibitem [{\citenamefont {Bernhard}\ \emph {et~al.}(2016)\citenamefont
  {Bernhard}, \citenamefont {Moreland}, \citenamefont {Bass}, \citenamefont
  {Liu},\ and\ \citenamefont {Heinz}}]{Bernhard:2016tnd}%
  \BibitemOpen
  \bibfield  {author} {\bibinfo {author} {\bibfnamefont {J.~E.}\ \bibnamefont
  {Bernhard}}, \bibinfo {author} {\bibfnamefont {J.~S.}\ \bibnamefont
  {Moreland}}, \bibinfo {author} {\bibfnamefont {S.~A.}\ \bibnamefont {Bass}},
  \bibinfo {author} {\bibfnamefont {J.}~\bibnamefont {Liu}}, \ and\ \bibinfo
  {author} {\bibfnamefont {U.}~\bibnamefont {Heinz}},\ }\href {\doibase
  10.1103/PhysRevC.94.024907} {\bibfield  {journal} {\bibinfo  {journal} {Phys.
  Rev.}\ }\textbf {\bibinfo {volume} {C94}},\ \bibinfo {pages} {024907}
  (\bibinfo {year} {2016})},\ \Eprint {http://arxiv.org/abs/1605.03954}
  {arXiv:1605.03954 [nucl-th]} \BibitemShut {NoStop}%
\bibitem [{Note4()}]{Note4}%
  \BibitemOpen
  \bibinfo {note} {See \cite {Aamodt:2010pb,Adam:2015ptt} and references
  therein.}\BibitemShut {Stop}%
\bibitem [{\citenamefont {Dusling}\ \emph {et~al.}(2016)\citenamefont
  {Dusling}, \citenamefont {Li},\ and\ \citenamefont
  {Schenke}}]{Dusling:2015gta}%
  \BibitemOpen
  \bibfield  {author} {\bibinfo {author} {\bibfnamefont {K.}~\bibnamefont
  {Dusling}}, \bibinfo {author} {\bibfnamefont {W.}~\bibnamefont {Li}}, \ and\
  \bibinfo {author} {\bibfnamefont {B.}~\bibnamefont {Schenke}},\ }\href
  {\doibase 10.1142/S0218301316300022} {\bibfield  {journal} {\bibinfo
  {journal} {Int. J. Mod. Phys.}\ }\textbf {\bibinfo {volume} {E25}},\ \bibinfo
  {pages} {1630002} (\bibinfo {year} {2016})},\ \Eprint
  {http://arxiv.org/abs/1509.07939} {arXiv:1509.07939 [nucl-ex]} \BibitemShut
  {NoStop}%
\bibitem [{Note5()}]{Note5}%
  \BibitemOpen
  \bibinfo {note} {If this is not the case, the system rapidly expands in all
  three dimensions, resulting in a rapid decrease of the energy density such
  that the system freezes out before reaching the hydrodynamic
  stage.}\BibitemShut {Stop}%
\bibitem [{\citenamefont {Muller}\ and\ \citenamefont
  {Rajagopal}(2005)}]{Muller:2005en}%
  \BibitemOpen
  \bibfield  {author} {\bibinfo {author} {\bibfnamefont {B.}~\bibnamefont
  {Muller}}\ and\ \bibinfo {author} {\bibfnamefont {K.}~\bibnamefont
  {Rajagopal}},\ }\bibfield  {booktitle} {\emph {\bibinfo {booktitle}
  {{Proceedings, 1st International Conference on Hard and Electromagnetic
  Probes of High-Energy Nuclear Collisions (Hard Probes 2004): Ericeira,
  Portugal, November 4-10, 2004}}},\ }\href {\doibase
  10.1140/epjc/s2005-02256-3} {\bibfield  {journal} {\bibinfo  {journal} {Eur.
  Phys. J.}\ }\textbf {\bibinfo {volume} {C43}},\ \bibinfo {pages} {15}
  (\bibinfo {year} {2005})},\ \Eprint {http://arxiv.org/abs/hep-ph/0502174}
  {arXiv:hep-ph/0502174 [hep-ph]} \BibitemShut {NoStop}%
\bibitem [{\citenamefont {Hanus}(2018)}]{Hanusthesis}%
  \BibitemOpen
  \bibfield  {author} {\bibinfo {author} {\bibfnamefont {P.}~\bibnamefont
  {Hanus}},\ }\emph {\bibinfo {title} {Entropy in Pb-Pb Collisions at the
  LHC}},\ \href
  {http://www.physi.uni-heidelberg.de//Publications/Bachelor_Thesis_Patrick_Hanus_2018.pdf}
  {\bibinfo {type} {Bachelor's thesis}},\ \bibinfo  {school} {University of
  Heidelberg} (\bibinfo {year} {2018}),\ \bibinfo {note} {supervisor Prof.
  Klaus Reygers}\BibitemShut {NoStop}%
\bibitem [{\citenamefont {Başar}\ and\ \citenamefont
  {Teaney}(2014)}]{Basar:2013hea}%
  \BibitemOpen
  \bibfield  {author} {\bibinfo {author} {\bibfnamefont {G.}~\bibnamefont
  {Başar}}\ and\ \bibinfo {author} {\bibfnamefont {D.}~\bibnamefont
  {Teaney}},\ }\href {\doibase 10.1103/PhysRevC.90.054903} {\bibfield
  {journal} {\bibinfo  {journal} {Phys. Rev.}\ }\textbf {\bibinfo {volume}
  {C90}},\ \bibinfo {pages} {054903} (\bibinfo {year} {2014})},\ \Eprint
  {http://arxiv.org/abs/1312.6770} {arXiv:1312.6770 [nucl-th]} \BibitemShut
  {NoStop}%
\bibitem [{\citenamefont {Aamodt}\ \emph
  {et~al.}(2010{\natexlab{a}})\citenamefont {Aamodt} \emph
  {et~al.}}]{Aamodt:2010pp}%
  \BibitemOpen
  \bibfield  {author} {\bibinfo {author} {\bibfnamefont {K.}~\bibnamefont
  {Aamodt}} \emph {et~al.} (\bibinfo {collaboration} {ALICE}),\ }\href
  {\doibase 10.1140/epjc/s10052-010-1350-2} {\bibfield  {journal} {\bibinfo
  {journal} {Eur. Phys. J.}\ }\textbf {\bibinfo {volume} {C68}},\ \bibinfo
  {pages} {345} (\bibinfo {year} {2010}{\natexlab{a}})},\ \Eprint
  {http://arxiv.org/abs/1004.3514} {arXiv:1004.3514 [hep-ex]} \BibitemShut
  {NoStop}%
\bibitem [{\citenamefont {Abelev}\ \emph {et~al.}(2013)\citenamefont {Abelev}
  \emph {et~al.}}]{ALICE:2012xs}%
  \BibitemOpen
  \bibfield  {author} {\bibinfo {author} {\bibfnamefont {B.}~\bibnamefont
  {Abelev}} \emph {et~al.} (\bibinfo {collaboration} {ALICE}),\ }\href
  {\doibase 10.1103/PhysRevLett.110.032301} {\bibfield  {journal} {\bibinfo
  {journal} {Phys. Rev. Lett.}\ }\textbf {\bibinfo {volume} {110}},\ \bibinfo
  {pages} {032301} (\bibinfo {year} {2013})},\ \Eprint
  {http://arxiv.org/abs/1210.3615} {arXiv:1210.3615 [nucl-ex]} \BibitemShut
  {NoStop}%
\bibitem [{\citenamefont {Aamodt}\ \emph
  {et~al.}(2010{\natexlab{b}})\citenamefont {Aamodt} \emph
  {et~al.}}]{Aamodt:2010pb}%
  \BibitemOpen
  \bibfield  {author} {\bibinfo {author} {\bibfnamefont {K.}~\bibnamefont
  {Aamodt}} \emph {et~al.} (\bibinfo {collaboration} {ALICE}),\ }\href
  {\doibase 10.1103/PhysRevLett.105.252301} {\bibfield  {journal} {\bibinfo
  {journal} {Phys. Rev. Lett.}\ }\textbf {\bibinfo {volume} {105}},\ \bibinfo
  {pages} {252301} (\bibinfo {year} {2010}{\natexlab{b}})},\ \Eprint
  {http://arxiv.org/abs/1011.3916} {arXiv:1011.3916 [nucl-ex]} \BibitemShut
  {NoStop}%
\bibitem [{\citenamefont {Adam}\ \emph {et~al.}(2016)\citenamefont {Adam} \emph
  {et~al.}}]{Adam:2015ptt}%
  \BibitemOpen
  \bibfield  {author} {\bibinfo {author} {\bibfnamefont {J.}~\bibnamefont
  {Adam}} \emph {et~al.} (\bibinfo {collaboration} {ALICE}),\ }\href {\doibase
  10.1103/PhysRevLett.116.222302} {\bibfield  {journal} {\bibinfo  {journal}
  {Phys. Rev. Lett.}\ }\textbf {\bibinfo {volume} {116}},\ \bibinfo {pages}
  {222302} (\bibinfo {year} {2016})},\ \Eprint
  {http://arxiv.org/abs/1512.06104} {arXiv:1512.06104 [nucl-ex]} \BibitemShut
  {NoStop}%
\end{thebibliography}%

\end{document}